\documentclass[aps,twocolumn]{revtex4}

\usepackage{graphics}
\usepackage{epsfig}
\usepackage{bm}
\usepackage{amsmath,amssymb}

\def\ov#1{\overline{#1}}

\def\wt#1{\widetilde{#1}}
\def\vb#1{\mbox{\boldmath$#1$}}
\def\pd#1#2{\frac{\partial #1}{\partial #2}}
\def\fd#1#2{\frac{\delta #1}{\delta #2}}
\def\td#1{\frac{\partial #1}{\partial t}}
\def\rtd#1{\frac{\partial_\epsilon #1}{\partial t}}
\def\wh#1{\widehat{#1}}
\def\bdot{\,\vb{\cdot}\,}
\def\btimes{\,\vb{\times}\,}

\def\bhat{\wh{{\sf b}}}
\def\cal#1{\mathcal{#1}}

\def\bhat{\wh{{\sf b}}}
\def \calF{{\mathcal F}}
\def\calH{{\mathcal H}}
\def\calS{{\mathcal S}}

\newcommand{\bc}{\begin{center}}
\newcommand{\ec}{\end{center}}
\newcommand{\bt}{\begin{tabbing}}
\newcommand{\et}{\end{tabbing}}
\newcommand{\be}{\begin{equation}}
\newcommand{\ee}{\end{equation}}
\newcommand{\ben}{\begin{eqnarray}}
\newcommand{\een}{\end{eqnarray}}

\newcommand{\bX}{\mathbf{X}}
\newcommand{\pa}{\partial}
\newcommand{\hb}{\wh{\sf b}}

\newcommand{\sg}{\mathcal{S}}

\newcommand{\ha}{\wh{\sf a}}
\newcommand{\hc}{\wh{\sf c}}
\newcommand{\oed}{\mathcal{O}(\epsilon^2)}

\begin{document}

\title{Hamiltonian formulation of reduced Vlasov-Maxwell equations}

\author{C. Chandre$^1$, A. J. Brizard$^{1,2}$, E. Tassi$^1$}
\affiliation{$^1$ Centre de Physique Th\'{e}orique -- CNRS/Aix-Marseille Universit\'{e} (UMR 6207), Campus de Luminy, case 907, 13009 Marseille, France\\$^2$ Department of Chemistry and Physics, Saint Michael's College, Colchester, VT 05439, USA}

\begin{abstract}
The Hamiltonian formulation of the reduced Vlasov-Maxwell equations is expressed in terms of the macroscopic fields ${\bf D}$ and ${\bf H}$. These macroscopic fields are themselves expressed in terms of the functional Lie-derivative $\pounds_{{\cal S}}$ generated by the functional ${\cal S}$, where
$\pounds_{{\cal S}}{\cal F} \equiv [{\cal S},\;{\cal F}]$ is expressed in terms of the Poisson bracket $[\;,\;]$ for the exact Vlasov-Maxwell equations. Hence, the polarization vector ${\bf P} \equiv ({\bf D} - {\bf E})/4\pi$ and the magnetization vector ${\bf M} \equiv ({\bf B} - {\bf H})/
4\pi$ are defined in terms of the expressions $4\pi\,{\bf P} \equiv \,[{\cal S},\;{\bf E}] + \cdots$ and $4\pi\,{\bf M} \equiv -[{\cal S},\;{\bf B}] + \cdots$, where lowest-order terms yield dipole contributions.
\end{abstract}

\begin{flushright}
October 30, 2012
\end{flushright}

%\pacs{}

\maketitle

\section{Introduction}

Averaging dynamical equations over small spatial and temporal scales allows a simplified description and an easier numerical integration of these equations in order to identify the basic mechanisms at play. This averaging is traditionally performed by defining macroscopic quantities defined as integrals of the corresponding microscopic quantities over small volumes and a small time interval. The macroscopic equations that are the dynamical equations for the macroscopic quantities are also obtained by computing the integrals of the microscopic equations. Once obtained, these macroscopic equations are better suited to the investigation of the underlying processes at the space and time scales of interest. In general, this procedure suffers from three drawbacks: The first drawback is that the basic structure of the microscopic equations (e.g., Hamiltonian structure) might be lost in the integration process. The second drawback is that this procedure is associated with a loss of information, which means that once  macroscopic quantities are computed (e.g., numerically), there is no way to go back to the microscopic ones. The third drawback involves the so-called closure problem: one may obtain a hierarchy of macroscopic equations that cannot be closed unless additional assumptions are included in the model.

Canonical transformations allow one to perform such averages by offering a way to eliminate the dependency over fast variables. They have the significant advantage of not modifying the expression of the Poisson bracket. Another advantage is that the information on the microscopic scales is encapsulated in the change of coordinates, and therefore can be recovered at any step in the reduction procedure, provided that the transformation is (locally) invertible. Such reduction procedures are particularly useful in plasma physics.

In magnetized plasma physics, where the high number of particles prevents a direct integration of the dynamics, the only hope is to rely on averaged equations. This is the purpose of guiding-center theory or gyrokinetic theory \cite{BH_07,CB_09}. As a consequence of the elimination of the fast time scales, associated to the rapid motion of particles around the magnetic field, in the Vlasov equation, polarization and magnetization effects are introduced into the Maxwell equations.

In this article, we propose a method for the direct Hamiltonian reduction of the Vlasov-Maxwell equations. The reduced equations are expressed in terms of the macroscopic fields 
\begin{eqnarray}
{\bf D} & \equiv & {\bf E} \;+\; 4\pi\,{\bf P}, \label{eq:DEP} \\
{\bf H} & \equiv & {\bf B} \;-\; 4\pi\,{\bf M}, \label{eq:HBM}
\end{eqnarray}
which are obtained from the Vlasov-Maxwell Poisson bracket by the action of Lie transformations defined on a functional space. As an outcome of this procedure, we obtain expressions for the polarization ${\bf P}$ and magnetization ${\bf M}$ vectors. More precisely, we show how to define the fields 
${\bf D}$, ${\bf H}$ and $F$ by a reduction process, so that Eqs.~(\ref{eq:M_eq}) and (\ref{eq:E_div}) are mapped into the reduced Maxwell equations
\cite{Jackson,Schwinger,Brizard_08}
\begin{eqnarray}
\pd{{\bf D}}{t} & = & c\,\nabla\btimes{\bf H} \;-\; 4\pi\,{\bf J}_{\rm R}, \label{eq:D_t} \\
\nabla\bdot{\bf D} & = & 4\pi\,{\rho}_{\rm R}, \label{eq:D_div}
\end{eqnarray}
where ${\rho}_{\rm R}$ and ${\bf J}_{\rm R}$ are the reduced charge and current densities expressed in terms of moments of the reduced Vlasov distribution $F$. The remaining Maxwell equations
\begin{eqnarray}
\pd{\bf B}{t} & = & -\;c\,\nabla\btimes{\bf E}, \label{eq:B_t} \\
\nabla\bdot{\bf B} & = & 0 \label{eq:div_B}
\end{eqnarray}
are left unchanged by the functional transformation. The main advantage of the Hamiltonian reduction is that we preserve the Hamiltonian structure of the Vlasov-Maxwell equations at each step of the process.

In Sec.~\ref{sec:lie} we briefly review Lie transforms in the context of noncanonical and infinite dimensional Hamiltonian systems. In Sec.~\ref{sec:ext} we show that some of the results obtained for particles are lifted to the Poisson algebra of functionals. In Sec.~\ref{sec:magn} we apply Lie transforms to the Vlasov-Maxwell equations to derive the magnetization and polarization effects in magnetized plasmas. In particular we show that when the Lie transformation at the particle level does not depend on the electric field, the magnetic field is unchanged (no magnetization effects) and the reduced distribution is identical to the one obtained in the case of external fields. 

\section{Functional Lie transforms}
\label{sec:lie}

The idea of dynamical reduction that we present here relies on the existence of a small parameter, $\epsilon$, and consists of identifying, for a given dynamical system, another system (which we refer to as reduced system) ``close'' to the original one, in the sense that the two systems become identical when $\epsilon \rightarrow 0$. The reduced system possesses the desired properties, such as, for instance the existence of an invariant sub-space, so that the dynamics, up to some given order in $\epsilon$, is independent on a variable or on a parameter. 

Because our goal is to apply the method to Hamiltonian systems, such as the Vlasov equation, a convenient mapping which permits to find the reduced dynamics, is the Lie transform. Indeed the latter possesses the useful properties of being invertible, close to the identity for $\epsilon \rightarrow 
0$, and of being a Lie morphism \cite{Dep69}. Because the systems of interest here are infinite-dimensional, the dynamical variables being functions of coordinates $\mathbf{z}$ playing the role of labels, we make use of a Lie transform acting on {\it functionals} of the dynamical variables.   
Indeed, we consider a Poisson algebra consisting of functionals of some fields. We denote $\phi_i({\bf z})$ the dynamical field variables. The Poisson bracket is denoted $[\cdot,\cdot]$, and the dynamics of any functional $\calF[\phi_1({\bf z}),\ldots,\phi_N({\bf z})]$ is given by the Hamiltonian 
$\calH[\phi_1({\bf z}),\ldots,\phi_N({\bf z})]$:
\begin{equation}
\td{\calF} \;=\; [\calF,\calH].
\label{eq:func_dot_F}
\end{equation}
The Poisson bracket satisfies the antisymmetry property $[{\cal G}, {\cal F}] = -\,[{\cal F}, {\cal G}]$, the Leibnitz rule
$[{\cal F}\,{\cal G}, {\cal K}] = {\cal F}\,[{\cal G}, {\cal K}] + [{\cal F}, {\cal K}]\,{\cal G}$, and the Jacobi identity
\begin{equation}
\left[ {\cal F},\frac{}{} [{\cal G},\; {\cal K}] \right] \;+\; \left[ {\cal G},\frac{}{} [{\cal K},\; {\cal F}] \right] \;+\; \left[ {\cal K},
\frac{}{} [{\cal F},\; {\cal G}] \right] \;=\; 0,
\label{eq:Jacobi}
\end{equation}
where $({\cal F}, {\cal G}, {\cal K})$ are arbitrary functionals of the fields $\phi_i({\bf z})$.

We consider a given functional ${\cal S}$ of the fields $\phi_i$, and consider the {\it functional} Lie transform generated by this functional acting on any other functional ${\cal F}$ as
\begin{equation}
\label{eq:Lie}
{\rm e}^{-\epsilon \pounds_{{\cal S}}}{\cal F}={\cal F}-\epsilon[{\cal S},\; {\cal F}]+\frac{\epsilon^2}{2}\left[{\cal S},\; [{\cal S},\; {\cal F}]
\right]+\cdots.
\end{equation}
The reduced functionals are defined by
\begin{equation} \label{funcred}
\ov{\cal F}={\rm e}^{-\epsilon\pounds_{{\cal S}}}{\cal F}.
\end{equation}
Here there are two dynamics to be distinguished, even though they are generated using the same Poisson bracket $[\cdot,\cdot]$: the original dynamics generated by the Hamiltonian $\calH$ and the reduced dynamics generated by the reduced Hamiltonian $\ov{\calH}$. 

For the dynamical equations, we consider the reduced evolution operator defined by
\begin{equation}
\rtd{\ov{\cal F}}\equiv \left({\rm e}^{-\epsilon \pounds_{{\cal S}}} \frac{d}{d t}{\rm e}^{\epsilon \pounds_{{\cal S}}} \right)\ov{\cal F}={\rm e}^{-\epsilon\pounds_{{\cal S}}} \left[{\rm e}^{\epsilon\pounds_{{\cal S}}}\ov{\cal F},\frac{}{} {\rm e}^{\epsilon\pounds_{{\cal S}}}\ov{\cal H} \right],
\end{equation}
From the property
\begin{equation}
\label{lieexp}
\left[{\rm e}^{\epsilon\pounds_{{\cal S}}}{\cal F},{\rm e}^{\epsilon\pounds_{{\cal S}}}{\cal G}\right]= {\rm e}^{\epsilon\pounds_{{\cal S}}}\left[
{\cal F},\frac{}{}{\cal G}\right],\end{equation}
for any functionals ${\cal F}$ and ${\cal G}$, we deduce that Eq.~\eqref{eq:func_dot_F} becomes
\begin{equation}
\rtd{\ov{\cal F}} \;=\; [\ov{\cal F},\ov{\cal H}],
\label{eq:func_dot_F_eps}
\end{equation}
where the new Hamiltonian is defined by
\begin{equation}
\ov{\calH} \;=\; {\rm e}^{-\epsilon\pounds_{\cal S}}\calH \;=\; \calH -\epsilon\td{\cal S}+\frac{\epsilon^2}{2}\left[{\cal S},\td{\cal S} \right]
+\cdots,
\label{eq:ov_Ham}
\end{equation}
with $\partial{\cal S}/{\partial t}$ viewed as a notation for $[{\cal S},{\cal H}]$. 

Since the present framework is that of noncanonical Hamiltonian systems, there exists in general a special class of Casimir invariants, which Poisson-commute with any observable $\calF$. In other words, a Casimir invariant ${\cal C}$ satisfies $[{\cal C},\calF ]=0$ for any functional $\calF$. These Casimir invariants are unchanged by functional Lie-transforms \eqref{funcred}, i.e., $\ov{\cal C}={\rm e}^{-\epsilon\pounds_{\cal S}}{\cal C}={\cal C}$ since $[{\cal S},{\cal C}]=0$.  

The reduced dynamics is linked to the dynamics of reduced field variables. First we define the reduced variables~: 
\begin{equation}
\psi_i({\bf z})={\rm e}^{\epsilon \pounds_{\cal S}} \phi_i({\bf z}).
\label{eq:psi_def}
\end{equation}
Given the scalar invariance $\bar{\cal F}(\psi_i)= {\cal F}(\phi_i)$, we recover Eq.~(\ref{funcred}) when expressed in the original field variables $\phi_i$. The dynamics of these reduced variables is the same as the reduced dynamics of the original variables, i.e.,
\begin{equation}
\label{eq:phipsi}
{\rm e}^{-\epsilon\pounds_{\cal S}} \td{\psi_i}=\rtd{\phi_i},
\end{equation}
where the left hand side of the equation contains the Lie transform in order to express the dynamics of the reduced field variables $\psi_i$ into the old field variables $\phi_i$. The above expression can be rewritten as
\begin{equation}
{\rm e}^{-\epsilon\pounds_{\cal S}} [\psi_i,\bar{\cal H}]_\epsilon =[\phi_i,\bar{\cal H}], 
\label{eq:id_psi}
\end{equation}
which is obvious according to the property~(\ref{lieexp}) since $[\psi_i,\bar{\cal H}]_\epsilon=[{\rm e}^{\epsilon \pounds_{\cal S}} \phi_i, {\cal H}]$. Here the bracket $[\cdot,\cdot]_\epsilon$ denotes the Poisson bracket $[\cdot,\cdot]$ expressed in the new field variables $\psi_i$. 

\section{Case of external electromagnetic fields}
\label{sec:ext}

We consider the Vlasov equation in time-dependent external electromagnetic fields, for the particle density $f({\bf x},{\bf v},t)$ in phase space~:
\begin{equation}
\pd{f}{t} \;=\; -{\bf v}\bdot \nabla f-\frac{e}{m}\left({\bf E}+\frac{\bf v}{c}\times {\bf B}\right)\bdot\pd{f}{\bf v}.
\label{eq:Vlasov_eq}
\end{equation}
The dynamics is generated by the Hamiltonian 
\begin{equation}
{\cal H}(f,w,\tau) \equiv \int \frac{m}{2}|{\bf v}|^{2} f d^6z \;-\; w,
\label{eq:H_ext}
\end{equation} 
where ${\bf z}=({\bf x},{\bf v})$ and the extra variable $w$ is due to the explicit time dependence of the fields ($\tau$ is its canonically conjugate variable). The Poisson bracket between two functionals in the extended phase space $(f; w,\tau)$ is given by
\begin{eqnarray}
\left[{\cal F},{\cal G}\right] &=&  \int f \left\{ \fd{{\cal F}}{f}, \fd{{\cal G}}{f} \right\} d^{6}z \nonumber \\
 &  &+\frac{e}{m} \int   {\bf E}\bdot \frac{\partial f}{\partial {\bf v}} \left( \fd{{\cal F}}{f}\frac{\partial {\cal G}}{\partial w}- \frac{\partial {\cal F}}{\partial w}\fd{{\cal G}}{f}  \right)d^{6}z  \nonumber \\
 &&+\frac{\partial {\cal F}}{\partial w}\frac{\partial {\cal G}}{\partial \tau}-\frac{\partial {\cal F}}{\partial \tau}
\frac{\partial {\cal G}}{\partial w},
\end{eqnarray}
where the bracket $\{\cdot,\cdot \}$ is given by
\begin{eqnarray}
\label{eq:pb}
\{ f,\; g\} & \equiv & \frac{1}{m} \left( \nabla f\bdot\pd{g}{{\bf v}} \;-\; \pd{f}{{\bf v}}\bdot\nabla g \right) \nonumber \\
 &  &+\; \frac{e\,{\bf B}}{m^{2}c}\bdot\left(\pd{f}{{\bf v}}\times\pd{g}{{\bf v}}\right),
\end{eqnarray}
that is to say that it is the noncanonical Poisson bracket for charged particles in a magnetic field, where the last term appears as a result of the use of the particle velocity ${\bf v}$ instead of the canonical momentum. 

\subsection{Functional Lie-transform}

We perform a Lie transform generated by a functional ${\cal S}(f,w,\tau)$, so that the functionals ${\cal F}$ are changed into $\ov{\cal F}=\exp (-\epsilon \pounds_{\cal S}) {\cal F}$. The reduced variables $(F,\ov{w},\ov{\tau})$ are transformed into
\begin{equation}
F={\rm e}^{\epsilon\pounds_{{\cal S}}} f, \qquad \bar{w}={\rm e}^{\epsilon\pounds_{{\cal S}}} w, \qquad \bar{\tau}={\rm e}^{\epsilon\pounds_{{\cal S}}} \tau.
\label{eq:FWTau}
\end{equation}
We impose the natural condition that the time variable is unchanged ($\ov{\tau}=\tau$), which implies that $\partial {\cal S}/\partial w=0$. The distribution function is mapped into
\begin{eqnarray}
F & \equiv & e^{\epsilon\pounds_{\calS}}\,f \;=\; f \;+\; \epsilon\,\left[ \calS,\frac{}{} f\right] + \cdots \nonumber \\
 & = & f \;-\; \epsilon\,\left\{\sigma,\frac{}{} f \right\} + \cdots,
\label{eq:Ff_def}
\end{eqnarray}
where the second expression shows the standard push-forward operation on the function $f({\bf z})$ generated by the function $\sigma \equiv \delta{\cal S}/{\delta f}$ and associated with the Poisson bracket $\{\cdot,\cdot\}$. 

For the reduction procedure, it is now more convenient to work at the level of the coordinates ${\bf z}$ of the particles. This is what is conventionally done in gyrokinetic theory \cite{BH_07}. More specifically, we present below an application of the proposed method to the reduction in the guiding center theory in the case of a static magnetic field and in the absence of electric field. In this example the reduction procedure consists in obtaining, under the assumption of high cyclotron frequency, a Vlasov system describing, up to a given order, the evolution of a guiding center distribution function, which represents the density in a phase space which is ``smaller'' than the initial one. More precisely, the reduction consists in eliminating the dependence on the particle gyration angle, which evolves on time scales much shorter than those of interest. The reduction procedure described in this example is based, to a great extent, on the analysis that was carried out by Littlejohn, at the level of particle dynamics in Refs. \cite{Lit79,Lit81}. 

\subsection{Guiding-center Problem}

In a nutshell, Littlejohn \cite{Lit81} derived a sequence of (in general noncanonical) transformations that, starting from the canonical coordinates used to describe the motion of a charged particle, led to a new set of coordinates, in terms of which, the corresponding dynamics is independent of the gyration angle, up to terms of the desired order in the small parameter $\epsilon$, which represents the ratio of the gyration period of the particle (with mass and charge set to unity) with respect to the characteristic time scale of the guiding center dynamics. For the sake of our example, we only consider the last step in the procedure which performs the gyroangle average. We denote the new coordinates with $(\mathbf{X},U,J,\theta)$ and their relation with the original particle coordinates will be given below.
 
The starting point for our example is the Hamiltonian functional (up to first order in $\epsilon$)
\begin{eqnarray} 
\calH(F)&=&\int d^6 Z F\;\left( H_{0}  \;+\frac{}{} \epsilon\,H_{1}\right),
\label{h}
\end{eqnarray}
where $H_{0} = \frac{1}{2}\,U^2 + J\,B$ denotes the unperturbed Hamiltonian, the first-order Hamiltonian is
\begin{eqnarray} 
H_{1} &=& \vb{\rho}_{0}\bdot\left(U^2\;\vb{\kappa} + \frac{2}{3}\,J\nabla B\right) \nonumber \\
 &  &+ \frac{1}{2}\,JU \left( \bhat\bdot\nabla\btimes\bhat - 2\alpha_{1}\right)
\label{eq:H1}
\end{eqnarray}
and the bracket (note that ${\bf E} = 0$)
\be  \label{par}
[{\cal F},{\cal G}]=\int d^6 Z\;  F\;\left\{{\cal F}_{F},\frac{}{}{\cal G}_{F}\right\}_{gc},
\ee
with subscripts on functionals indicating functional derivatives. In Eq.~(\ref{eq:H1}), $\vb{\rho}_{0} \equiv (2J/B)^{1/2}\,\ha$ denotes the lowest-order gyroradius vector, $\vb{\kappa} \equiv \bhat\bdot\nabla\bhat$ denotes the magnetic curvature, and $\alpha_{1} \equiv -\frac{1}{2}\,(\hc\bdot\nabla\bhat\bdot\ha + \ha\bdot\nabla\bhat\bdot\hc)$, with $\hc$ defined such that $\ha=\hb\times\hc$. The gyration angle of the particle motion is indicated with $\theta$. An alternative orthonormal basis $(\hat{1},\hat{2},\hb)$ with $\ha=\hat{1}\cos\theta -\hat{2}\sin\theta $, can be defined and through which the dependence of $\ha$ and $\hc$ on $\theta$ can be made explicit.

The distribution function $F(\bX,U,J,\theta)$ is the density in phase space identified by the coordinates (defined up to first order in $\epsilon$)
\ben
\bX &=& \mathbf{x}-\epsilon \;\vb{\rho}_{0},\\
U &=& u -\epsilon \left[u\;\vb{\kappa}\bdot\vb{\rho}_{0} - J_{0}\left( \frac{1}{2}\,\bhat\bdot\nabla\btimes\bhat + \alpha_{1} \right)\right],\\
J &=& J_{0} \left[ 1 -\epsilon\,\left(\frac{u}{B}\,\bhat\bdot\nabla\btimes\bhat - \frac{1}{3}\,\vb{\rho}_{0}\bdot\nabla\ln B\right)\right],\\      
\theta &=& \zeta \;-\; \epsilon\,\vb{\rho}_{0}\bdot{\bf R},
\een  
where $J_{0} \equiv v_{\bot}^{2}/2B$, $\mathbf{x}$ is the particle position, $u=\mathbf{v}\bdot\hb$ and $v_{\perp}=\vert\mathbf{v}\times \hb\vert$ its parallel and perpendicular velocity, respectively, and $\zeta$ is the physical (instantaneous) gyration angle, and ${\bf R} \equiv \nabla\wh{\sf 1}\bdot
\wh{\sf 2}$ denotes the gyrogauge vector. The inner bracket in the bracket (\ref{par}) is
\begin{eqnarray}
\{f ,g\}_{gc} & = & \epsilon^{-1}\left(\frac{\pa f}{\pa \theta}\frac{\pa g}{\pa J}-\frac{\pa f}{\pa J}\frac{\pa g}{\pa \theta}\right)\nonumber \\
 &  &+\; \left(\hb + \epsilon\;\frac{U}{B}\,\bhat\btimes\vb{\kappa}\right)\bdot\left(\frac{\pa f}{\pa \bX}\frac{\pa g}{\pa U}- \frac{\pa g}{\pa \bX}
\frac{\pa f}{\pa U}  \right) \nonumber \\
 &  &-\;\epsilon\frac{\hb}{B}\bdot\left(\frac{\pa f}{\pa \bX}\times\frac{\pa g}{\pa \bX}\right).
\label{eq:gc_PB}
\end{eqnarray}
Because the magnetic field is external and no electric field is present, the observables of this Hamiltonian system consist of the functionals ${\cal F}(F)$, which depend on the distribution function $F$ only. Notice that the Hamiltonian (\ref{h}) is of the form 
where $H_1$ depends explicitly on $\theta$. With the help of the functional Lie transform procedure described in Sec.~\ref{sec:lie}, we show how this dependence is removed at order $O(\epsilon)$ and pushed to terms of order $O(\epsilon^2)$. 

According to Eq.~(\ref{funcred}), the transformed Hamiltonian reads
\be  \label{trh}
\begin{split}
\bar{\cal H}(F)&= {\rm e}^{-\epsilon\pounds_{{\cal S}}}{\cal H}(F)={\cal H}(F)-\epsilon [\sg(F),{\cal H}(F)]+ \cdots\\
&=\int d^6 Z (H_0+ \epsilon (H_1-\{\sg_{F},H_0 \}_{gc}) + \cdots) F \\
\end{split}
\ee
The generating function $\cal S$ is determined in such a way that it cancels all terms explicitly dependent on the gyro-angle $\theta$ in $H_1$. 
Assuming that the transformed Hamiltonian functional admits an expansion as
\be  \label{bh}
\bar{\cal H}(F)=\int d^6 Z (\bar{H_0}+ \epsilon {\bar H_1})F + \oed
\ee
and comparing this last expression with (\ref{trh}), one obtains
\begin{eqnarray}
\bar{H_0}=H_0,\\
\bar{H_1}=H_1-\{\sg_{F},H_0\}_{gc}.
\end{eqnarray}
Bearing in mind that we want $\bar{H_1}$ to be independent of $\theta$, the first-order relation can be separated into $\theta$-dependent and $\theta$-independent parts:
\be  \label{c1}
\bar{H_1}=\langle H_1\rangle,
\ee
\be  \label{c2}
0= \wt{H_1}- \{\sg_{F},H_0\}_{gc},
\ee
where $\langle \cdot  \rangle$ denotes $\theta$-averaging and $\wt{H} \equiv H - \langle H\rangle$ denotes the $\theta$-dependent part of $H$ (it is assumed that $\langle \sg_{F}\rangle \equiv 0$). The condition~(\ref{c1}) tells us that
\be
\bar{H_1}\;=\; J\,U\;\left(\frac{1}{2}\;\bhat\bdot\nabla\btimes\bhat\right).
\ee
On the other hand, the condition (\ref{c2}) gives us an equation to be solved with respect to $\sg_{F}$. Its solution reads \cite{Lit81}
\begin{eqnarray} \label{sols}
\sg_{F} & = & \pd{\vb{\rho}_{0}}{\theta}\bdot\left(\frac{U^2}{B}\;\vb{\kappa} + \frac{2}{3}\,J\nabla \ln B\right) + \frac{JU}{B}\,\alpha_{2},
\end{eqnarray}
where $\alpha_{1} \equiv \partial\alpha_{2}/\partial\theta$. Therefore, the functional that generates the required Lie transform is 
\be   \nonumber
\sg(F)=\int d^6 Z F \sg_{F},
\ee
with $\sg_{F}$ given by Eq.~(\ref{sols}). 
The resulting Vlasov equation is then given by
\begin{equation}
\frac{\partial_{\epsilon} F}{\partial t} = [F,\bar{\cal H}(F)]=\{\bar{H_0}, F\}_{gc} + \epsilon \{\bar{H_1}, F\}_{gc}+\cdots.
\end{equation}
Because $\bar{H_0}$ and $\bar{H_1}$ are independent of $\theta$, we identify an invariant sub-algebra, consisting of the functionals ${\cal F}(\langle F\rangle)$ of $\theta$-independent distribution functions $\langle F\rangle$. 

Indeed, if the distribution function at $t=0$ is $\theta$-independent, then, evolving according to the dynamics generated by $[.,\bar{\cal H}_{gc}]$, it will remain $\theta$-independent at any time (since the Hamiltonian $\bar{\cal H}_{gc}$ is also an element of this sub-algebra since it can be rewritten as a functional of $\langle F\rangle$). This Lie sub-algebra realizes the dynamical reduction. According to (\ref{eq:phipsi}), the dynamics of the reduced distribution function $\bar{F} (\bX,U,J,\theta)={\rm e}^{\epsilon\pounds_{{\cal S}}}F (\bX,U,J,\theta)$, is given by
\be
\frac{\partial \bar{f}_{gc}}{\partial t}={\rm e}^{\epsilon\pounds_{{\cal S}}}\frac{\partial_{\epsilon} F}{\partial t}.
\ee
%

%We have thus shown with an explicit example how, with the help of the functional Lie transform, it is possible to obtain a reduced Vlasov equation, with a known Hamiltonian structure. In the example under consideration, the reduction made it possible to remove the influence on the dynamics of the fast variable corresponding to the particle gyration angle. Should one want to express the reduced dynamics in terms of Lie transformed variables $(\bar{f},\mathbf{D},\mathbf{H})=({\rm e}^{\epsilon \pounds_{{\cal S}}}f, {\rm e}^{\epsilon \pounds_{{\cal S}}} \mathbf{E}, {\rm e}^{\epsilon \pounds_{{\cal S}}} \mathbf{B})$ (note that, due to the scalar invariance, the Lie transform for the change of variables is the inverse of the Lie transform for changing the functionals) then the governing equation for a generic functional $\bar{\cal F}(\bar{f},\mathbf{D},\mathbf{H})$ would be
%%
%\begin{eqnarray}  \label{newv}
%\frac{\partial \bar{\cal F}(\bar{f},\mathbf{D},\mathbf{H})}{\partial t}=[ \bar{\cal F}(\bar{f},\mathbf{D},\mathbf{H}),\bar{\bar{H}}(\bar{f},\mathbf{D},\mathbf{H})]_{\epsilon}={\rm e}^{-\epsilon \pounds_{{\cal S}}}[ {\cal F}(\bar{f},\mathbf{D},\mathbf{H}),\bar{\cal H}(\bar{f},\mathbf{D},\mathbf{H})]_{\epsilon}=[{\cal F}(f,\mathbf{E},\mathbf{B}),{\cal H}(f,\mathbf{E},\mathbf{B})].
%\end{eqnarray}
%%
%Note also that the two overbars that appear in correspondence to $H$ are required to indicate that we are considering the dynamics generated by the reduced Hamiltonian as a functional for the reduced variables. 

\section{Reduced Maxwell-Vlasov equations} 
\label{sec:magn}
 
The next example we consider is the case of a collisionless plasma whose dynamics is described by the Vlasov-Maxwell equations. These equations give the dynamics of a distribution function $f({\bf x},{\bf v};t)$ of charged particles in phase space, together with an electric field ${\bf E}({\bf x};t)$ and magnetic field ${\bf B}({\bf x};t)$~:
\begin{eqnarray}
\pd{f}{t} & = & -\,{\bf v}\bdot\nabla f \;-\; \frac{e}{m} \left( {\bf E} + \frac{{\bf v}}{c}\btimes{\bf B}\right)\bdot\pd{f}{{\bf v}}, \label{eq:V_eq} \\
\pd{{\bf E}}{t} & = & c\,\nabla\btimes{\bf B} \;-\; 4\pi\,{\bf J}, \label{eq:M_eq} \\
\pd{{\bf B}}{t} & = & -\;c\,\nabla\btimes{\bf E}, \label{eq:F_eq}
\end{eqnarray}
where ${\bf J} \equiv e\int f {\bf v} d^{3}v$. The initial conditions of this dynamical system are constrained by the two remaining Maxwell's equations 
\begin{eqnarray}
\nabla\bdot{\bf E} & = & 4\pi\,\rho, \label{eq:E_div} \\
\nabla\bdot{\bf B} & = & 0, \label{eq:B_div}
\end{eqnarray}
where $\rho \equiv e\int  f  d^{3}v$, and a normalization condition
\begin{equation}
\int d^6z f = N.
\end{equation}
The Vlasov-Maxwell equations (\ref{eq:V_eq})-(\ref{eq:F_eq}) are expressed in Hamiltonian form using the Vlasov-Maxwell Poisson bracket~\cite{M,MW,B}
\begin{widetext}
\begin{eqnarray}
\left[{\cal F},\;{\cal G}\right] & = & \int\;f \left\{ \fd{{\cal F}}{f},\; \fd{{\cal G}}{f} \right\}\; d^{6}z \;+\; 4\pi\; \int d^{3}x \;
\fd{{\cal F}}{{\bf E}} \bdot \left( c\;\nabla\btimes\fd{{\cal G}}{{\bf B}} \;+\; \frac{e}{m} \int d^{3}v\;\fd{{\cal G}}{f}\;\pd{f}{{\bf v}} \right) \nonumber \\
 & &-\; 4\pi\; \int d^{3}x \;\fd{{\cal G}}{{\bf E}} \bdot \left( c\;\nabla\btimes\fd{{\cal F}}{{\bf B}} \;+\; \frac{e}{m} \int d^{3}v\;
\fd{{\cal F}}{f}\;\pd{f}{{\bf v}} \right),
\label{eq:MV_PB}
\end{eqnarray}
\end{widetext}
between two functionals $\cal F$ and $\cal G$ of the Vlasov-Maxwell fields $(f, {\bf E}, {\bf B})$, and the Hamiltonian is
\begin{equation}
{\cal H} \;\equiv\; \int \frac{m}{2}\,|{\bf v}|^{2}\;f\;d^6z \;+\; \int \frac{d^{3}x}{8\pi} \left( |{\bf E}|^{2} \;+\; |{\bf B}|^{2} \right),
\label{eq:H_def}
\end{equation}
The first term in Eq.~(\ref{eq:MV_PB}) involves the noncanonical Poisson bracket~(\ref{eq:pb}). 

We perform a Lie transform generated by ${\cal S}(f,{\bf E},{\bf B})$ changing the functional $\cal F$ as $\bar{\cal F}=\exp(-\epsilon\pounds_{\cal S}) 
{\cal F}$. First we start by defining the reduced field variables ${\bf D}$, ${\bf H}$ and $F$ as
\begin{equation}
\left( \begin{array}{c}
{\bf D} \\
{\bf H} \\
F
\end{array} \right) \;\equiv\; {\rm e}^{\epsilon\pounds_{{\cal S}}} \left( \begin{array}{c}
{\bf E} \\
{\bf B} \\
f
\end{array} \right).
\label{eq:EB_S}
\end{equation}
We notice the change of sign in the Lie transform which is a result of the scalar invariance $\bar{\cal F}(F, {\bf D},{\bf H})={\cal F}(f,{\bf E},{\bf B})$.
Also we notice that ${\bf D}$ and ${\bf H}$ only depend on ${\bf x}$ and $t$ since $\cal S$ has no explicit dependence on ${\bf z}=({\bf x},{\bf v})$.

We define the vectors ${\bf P}$ and ${\bf M}$ correspond to the polarization and magnetization effects as
\begin{equation}
\left( \begin{array}{c}
{\bf D} \\
\\
{\bf H}
\end{array} \right) \;\equiv\; \left( \begin{array}{c}
{\bf E} \;+\; 4\pi\,{\bf P} \\
\\
{\bf B} \;-\; 4\pi\,{\bf M}
\end{array} \right).
\label{eq:EBDH_def}
\end{equation}
Equivalently, these vectors can be defined as
\begin{eqnarray*}
&&{\bf P}=\frac{1}{4\pi}\left({\rm e}^{\epsilon\pounds_{{\cal S}}}-1 \right){\bf E}=\frac{1}{4\pi}[{\cal S},{\bf E}]+\cdots,\\
&&{\bf M}=\frac{1}{4\pi}\left(1-{\rm e}^{\epsilon\pounds_{{\cal S}}} \right){\bf B}=-\frac{1}{4\pi}[{\cal S},{\bf B}]+\cdots.
\end{eqnarray*}
We notice that $\nabla\bdot {\bf M}=0$ since ${\rm e}^{\epsilon\pounds_{{\cal S}}}$ commutes with $\nabla$.
From $\rho= \nabla\bdot {\bf E}/(4\pi)$, we have
$$
{\rho}_R=\nabla\bdot \left( {\rm e}^{\epsilon\pounds_{{\cal S}}} \frac{{\bf E}}{4\pi}\right)=\nabla\bdot \left( {\bf P} +\frac{{\bf E}}{4\pi}\right)=\rho+\nabla\bdot {\bf P},
$$
which corresponds to the standard equation for the reduced charge density (see Ref.~\cite{Brizard_08}). The first-order expressions for the polarization and magnetization are
\begin{eqnarray}
&& {\bf P}_1\equiv \frac{1}{4\pi}[{\cal S},{\bf E}] = -c\,\nabla\btimes\fd{{\cal S}}{{\bf B}} + e\int f {\bm\rho}_1 d^{3}v,
\label{eq:P_def}\\
&& {\bf M}_1\equiv -\frac{1}{4\pi}[{\cal S},{\bf B}] = -c\nabla\btimes\fd{{\cal S}}{{\bf E}}, \label{eq:M_def}
\end{eqnarray}
where the displacement ${\bm \rho}_1$ is defined by
$$
{\bm \rho}_1\equiv\frac{1}{m}\frac{\partial}{\partial {\bf v}}\left(\fd{\cal S}{f}\right),
$$
which agrees with Ref.~\cite{Brizard_08}. The first-order expansion for $F$ is given by
\begin{equation}
\label{eq:LSF}
F=f+\left\{f,\fd{\cal S}{f} \right\}+\frac{4\pi e}{m}\pd{f}{\bf v}\bdot \fd{\cal S}{\bf E}+\cdots.
\end{equation}
Section~\ref{sec:ext} gives a natural explanation of the two terms in Eq.~(\ref{eq:LSF})~: The first term of the right hand side of Eq.~(\ref{eq:LSF}) is associated with the action of the fields on the particles (like in the external field case), and the second one is intrinsic to the feedback from the particles to the fields.

Next, we look at the reduced dynamics of the field variables $(f,{\bf E},{\bf B})$ which is equivalent to looking at the dynamics of the reduced fields $(F,{\bf D},{\bf H})$ as we have seen in Sec.~\ref{sec:lie}. 
First we consider the reduced dynamics of $\bf B$ and rewrite it as~:
$$
\rtd{\bf B}=[{\bf B},\bar{\cal H}]={\rm e}^{-\epsilon\pounds_{{\cal S}}}\left([({\rm e}^{\epsilon\pounds_{{\cal S}}}-1){\bf B},\frac{}{}
{\cal H}]+[{\bf B},{\cal H}]\right).
$$
In the first term of the rightmost hand side, we recognize the time derivative of ${\bf M}$. By using Eqs.~(\ref{eq:phipsi}) and (\ref{eq:EBDH_def}), we recover the dynamics of the reduced field ${\bf H}$
$$
\td{\bf H}=-c\nabla \times {\bf D} +4\pi \left(c\nabla\times {\bf P}-\td{\bf M}\right)
$$  

A similar derivation is done for the dynamics of the reduced field ${\bf D}$ by using the reduced dynamics for ${\bf E}$ rewritten as
$$
\rtd{\bf E}=[{\bf E},\bar{\cal H}]={\rm e}^{-\epsilon\pounds_{{\cal S}}}\left([({\rm e}^{\epsilon\pounds_{{\cal S}}}-1){\bf E},\frac{}{}
{\cal H}]+[{\bf E},{\cal H}]\right).
$$
Using Eqs.~(\ref{eq:phipsi}) and (\ref{eq:EBDH_def}), the dynamics of ${\bf D}$ is given by
$$
\td{\bf D}=c\nabla \times {\bf H}-4\pi{\bf J}_R,
$$
where
$$
{\bf J}_R={\bf J}-c\nabla\times {\bf M}-\td{\bf P}. 
$$

In the case of external fields, a Lie transform in the functional space is associated with a phase space transformation at the particle level as it was seen in Sec.~\ref{sec:ext}. Below we consider the case where the Lie transform is generated by ${\cal S}$ given by
\begin{equation}
\label{eq:LieST}
{\cal S}(f,{\bf E},{\bf B})=\int f({\bf z}) \sigma({\bf z};{\bf E}({\bf x}),{\bf B}({\bf x})) d^6z,
\end{equation}
such that $\delta{\cal S}/\delta{f}=\sigma$. Furthermore we consider the case where the transformation done at the level of the particles only depends on ${\bf z}$ and on ${\bf B}$ as it is done in the guiding-center theory. In this case we have ${\cal S}_{\bf E}=0$, and hence, from Eq.~(\ref{eq:M_def}), the magnetization ${\bf M}$ is zero, which means that the magnetic field remains unchanged by the reduction procedure. 
In this particular case, we show that the distribution function is exactly the one obtained at the particle level by the Lie transform generated by $\sigma$ as in the case of external electromagnetic fields. First we notice that the right-most hand side of Eq.~(\ref{eq:LSF}) vanished. Second, we notice that the Poisson bracket~(\ref{eq:MV_PB}) between two functionals ${\cal F}$ and ${\cal G}$ of $f$ and ${\bf B}$ gives
$$
[{\cal F},{\cal G}]=\int f\left\{ {\cal F}_f,{\cal G}_f\right\} d^6 z.
$$
Therefore we show by induction that
$$
F={\rm e}^{-\epsilon \pounds_\sigma}f,
$$
where we recall that the Lie transformation is now generated with the Poisson bracket $\{\cdot,\cdot\}$ at the particle level. For instance the first order is already contained in Eq.~(\ref{eq:LSF}) and the second-order term
\begin{widetext}
$$
\left[{\cal S},\frac{}{}[{\cal S},f({\bf z}_0)]\right]=\int f({\bf z})\left\{ \sigma({\bf z};{\bf B}({\bf x})),\frac{\delta}{\delta f({\bf z})}\{f({\bf z}_0),\sigma({\bf z}_0;{\bf B}({\bf x}_0))\}\right\} d^6z.
$$
\end{widetext}
Given that 
$$
\frac{\delta}{\delta f ({\bf z})} \{f({\bf z}_0),\sigma({\bf z}_0;{\bf B}({\bf x}_0))\}=-\{\delta^6({\bf z}-{\bf z}_0),\sigma({\bf z};{\bf B}({\bf x}))\},
$$
by linearity and since the function $\sigma$ does not depend on $f$. Using two integrations by part (noticing that $\int d^6z \{f,g\}=0$ for all function $f$ and $g$ of ${\bf z}$), we obtain that
$$
\left[{\cal S},\frac{}{}[{\cal S},f({\bf z}_0)]\right]= \left\{\sigma({\bf z}_0;{\bf B}({\bf x}_0)),\frac{}{}\{\sigma({\bf z}_0;{\bf B}({\bf x}_0)),
f({\bf z}_0)\}\right\}.
$$
When the Lie transformations at the particle level depend on the electric field, the passage from the particle description to the functional level, the generating functional is no longer as straightforward to construct from the generating function at the particle level.  

\begin{acknowledgments}
CC and ET acknowledge useful discussions with the Nonlinear Dynamics team of the CPT. AJB acknowledges the warm hospitality of the CPT. This work was supported by the European Community under the contract of Association between Euratom, CEA and the French Research Federation for fusion studies. The views and opinions expressed herein do not necessarily reflect those of the European Commission. Financial support was also received from the Agence Nationale de la Recherche (ANR GYPSI). 
\end{acknowledgments}

\end{document}